%% file: main.tex
\documentclass[sigconf,nonacm]{acmart}

\usepackage{graphicx}
\usepackage{subfig}
\usepackage{xparse}
\usepackage{enumitem}
\usepackage{multirow}
\usepackage{tcolorbox}
\usepackage{xspace}
\usepackage{listings}
\usepackage{xcolor}
\usepackage{amsmath}
\usepackage{algorithm}
\usepackage{algpseudocode}
\usepackage{forest}
\usepackage[T1]{fontenc}
\usepackage[utf8]{inputenc}
\usepackage{appendix}

\DeclareMathOperator{\arccosh}{arccosh}
\DeclareMathOperator{\arcsinh}{arcsinh}

\DeclareMathOperator\erf{erf}

\tcbuselibrary{listings}
\newtcolorbox[auto counter, number within=section]{cuteconversationblue}[2][]{colback=blue!5!white, colframe=blue!75!white, colbacktitle=blue!10!white, coltitle=black, title=#2,#1}
\newtcolorbox[auto counter, number within=section]{cuteconversationred}[2][]{colback=red!5!white, colframe=red!75!white, colbacktitle=red!10!white, coltitle=black, title=#2,#1}
\newtcolorbox[auto counter, number within=section]{cuteconversationcyan}[2][]{colback=cyan!5!white, colframe=cyan!75!white, colbacktitle=cyan!10!white, coltitle=black, title=#2,#1}

\newcommand*\OR{\ |\ }

\newif\ifcomment\commentfalse
\commenttrue

\ifcomment
\newcommand\sadra[1]{\textcolor{red}{[\textbf{Sadra:} #1]}}
\newcommand\sepand[1]{\textcolor{blue}{[\textbf{Sepand:} #1]}}

\else 
\newcommand\sadra[1]{}
\newcommand\sepand[1]{}
\fi

\newcommand{\samila}{\textsf{Samila}\xspace}
\newcommand{\Samila}{\textsf{Samila}\xspace}

\title[Samila: A Generative Art Generator]{\samila: A Generative Art Generator}

\author{Sadra Sabouri$^*$}
\affiliation{ 
  \institution{Open Science Laboratory}
  \country{}
}
\email{sadra@openscilab.com}

\author{Sepand Haghighi$^*$}
\affiliation{ 
  \institution{Open Science Laboratory}
   \country{}
}
\email{sepand@openscilab.com}

\author{Elena Masrour}
\affiliation{ 
  \institution{Kansas State University}
   \country{}
}
\email{masrour@ksu.edu}

\begin{document}

\begin{abstract}
    Generative art merges creativity with computation, using algorithms to produce aesthetic works. This paper introduces \Samila\footnote{\url{https://github.com/sepandhaghighi/samila}}, a Python-based generative art library that employs mathematical functions and randomness to create visually compelling compositions. The system allows users to control the generation process through random seeds, function selections, and projection modes, enabling the exploration of randomness and artistic expression. By adjusting these parameters, artists can create diverse compositions that reflect intentionality and unpredictability. We demonstrate that \Samila's outputs are uniquely determined by two random generation seeds, making regeneration nearly impossible without both. Additionally, altering the point generation functions while preserving the seed produces artworks with distinct graphical characteristics, forming a visual family. \Samila serves as both a creative tool for artists and an educational resource for teaching mathematical and programming concepts. It also provides a platform for research in generative design and computational aesthetics. Future developments could include AI-driven generation and aesthetic evaluation metrics to enhance creative control and accessibility.
\end{abstract}

\maketitle
\def\thefootnote{*}\footnotetext{These authors contributed equally to this work}
\input{sections/introduction}
\input{sections/rw}
\input{sections/program-struct}
\input{sections/discussion}
\input{sections/future-works}
\input{sections/limit}

\input{sections/application}
\input{sections/conclusion}

\bibliography{main}
\bibliographystyle{acm}

\input{sections/appendix}

\end{document}

%% file: sections/introduction.tex
\begin{figure}
    \centering
    \includegraphics[width=\linewidth]{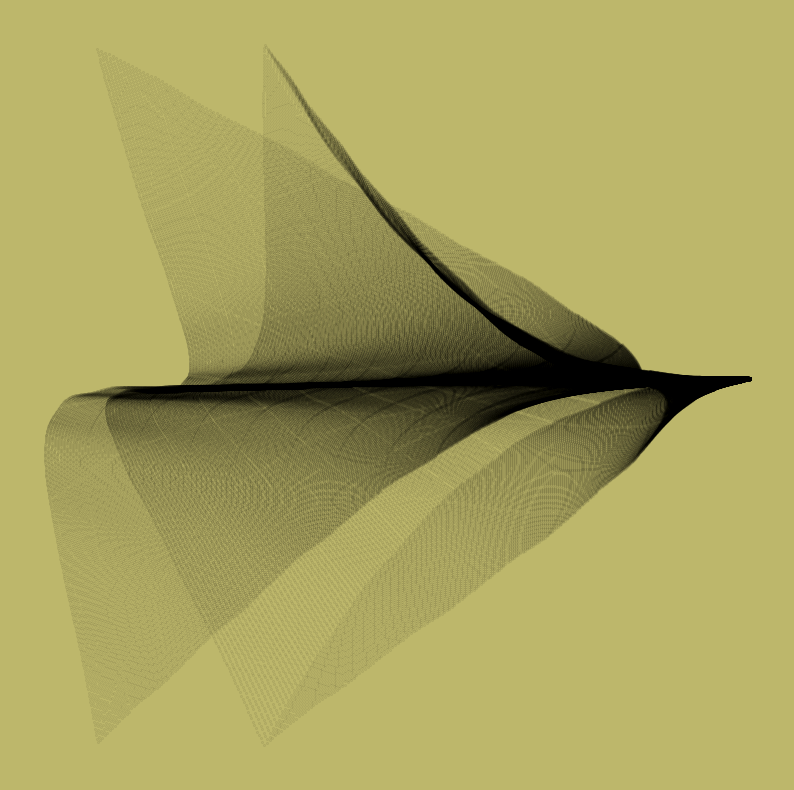}
    \caption{Example of \samila generative art.}
    \label{fig:example_fig}
\end{figure}

\section{Introduction}
Generative art represents a convergence of creativity and computation, using systems and algorithms to produce aesthetic expressions.
Galanter defined generative art as an art practice in which a system - rules, code, or machines - operates autonomously to shape the final work~\cite{galanter2003generative}.
This process-oriented approach emphasizes the intrinsic dynamism of the system.
Researchers highlight that generative art can shift attention from nouns to verbs, underscoring processes over outcomes, and embracing complexity across multiple scales and levels of emergence~\cite{galanter2016generative}.

Although the term ``generative art'' is typically associated with computer-based practices, the roots of generative methods lie in earlier 20\textsuperscript{th}-century art movements~\cite{kristiansen1968dada}.
Dadaists, for example, employed chance-based techniques in collage, while surrealists used automatic writing and drawing to tap into the subconscious~\cite{frey1936dada}.
Later, Abstract Expressionists such as Jackson Pollock adopted process-oriented painting methods, such as drizzling and pouring paint, with a measure of unpredictability.
His ``drip'' paintings introduced elements of randomness such as the interaction between fluid paint and gravity while maintaining enough intentional control to produce a balanced composition~\cite{rampley1996identity}.
This interplay of deliberate gestures and emergent complexity laid an important foundation for modern generative art, revealing how autonomous processes can lead to novel aesthetic outcomes.

With the advent of digital technology in the mid-20\textsuperscript{th} century, artists began writing computer programs to formalize creative rules and harness randomness algorithmically.
Pioneers like Frieder Nake, Vera Molnár, and Harold Cohen (AARON) demonstrated how logical structures and computer code could produce intricate, emergent visuals.
Their works illustrated the potential of algorithmic systems to transcend manual craft, echoing the same balance between intention and surprise observed in Pollock’s approach, yet now facilitated by computation~\cite{evans2023frieder,valyi2023vera,verdicchio2016role,dodds2019collecting}.

Contemporary generative art extends these historical trajectories into machine learning and artificial intelligence.
From Google DeepDream~\cite{mordvintsev2015deepdream} and the transfer of neural style to Generative Adversarial Networks (GANs)~\cite{goodfellow2020generative}, artists can train models on large datasets to produce output that mimics or re-imagines existing styles.
This approach introduces complex forms of authorship, where the ``artist'' designs training protocols and prompts, but the resulting images emerge from the learned representations of the model. Despite the novel tools, the enduring themes of chance, emergence, and system-driven creativity remain central.
Thus, AI-based generative art continues the lineage begun by Pollock's drips and the Dadaists' random collages, emphasizing process over final form and demonstrating how structured unpredictability can open new aesthetic frontiers.

Other generative systems rely on mathematical and algorithmic structures to shape their aesthetic explorations.
Gaussian Quadratic generative art, for instance, blends Gaussian distributions for horizontal placement with quadratic equations for vertical positioning, balancing order, and randomness in a structured interplay~\cite{benthall1972science}.
Assessing the aesthetic value of such outputs, however, remains an ongoing challenge.
While formulaic principles like the Golden Ratio~\cite{livio2008golden}, Fibonacci sequences, and Zipf’s law~\cite{powers-1998-applications} offer partial insights, information-theoretic approaches, such as those developed by Bense~\cite{bense1965projekte}, provide a more flexible yet still imperfect framework for evaluating generative art~\cite{galanter2012computational}.

Dorin et al. proposed a comprehensive framework for analyzing generative art, which includes four key components: \emph{Entities} involved in the creation, the \emph{Process} generating the art, its \emph{Environmental Interaction,} and the resulting \emph{Sensory Outcomes}~\cite{dorin2012framework}.
An example of this framework can be seen in Islamic star patterns from the ninth century, where points, lines, circles, and rhombuses served as the geometric entities for construction.
Their geometric boundaries defined the initialization and termination of shapes.
Although the exact processes underlying these patterns remain unknown, they likely followed general rules.
These patterns lacked environmental interaction but provided sensory outcomes as static visual works, often displayed on buildings or flat surfaces~\cite{dorin2012framework}.

Random generative art stands as a fascinating subset of the field, where systems are designed with rules that introduce an element of randomness to their outputs or use simulations of a random environment to generate art~\cite{schonlieb2013random}, resulting in unpredictable yet aesthetically pleasing results.
Such systems balance structure and chaos, often yielding outputs that evoke a sense of organic beauty.
For instance, Perlin noise, a gradient noise algorithm, has been widely used in creating natural-looking textures and terrains for generative art and computer graphics \citep{hart2001perlin}.
Similarly, L-systems, originally developed for modeling plant growth, employ stochastic rules to generate intricate, nature-inspired patterns \citep{hanan1992parametric}.
Another compelling example is the generative art framework built using Boid's rules for simulating flocking behavior in birds \citep{reynolds1987flocks}. This system modifies the proportion of total agents influencing each agent’s behavior based on its neighbors, resulting in dynamic, visually captivating patterns that mirror natural group behaviors.
These methods demonstrate how randomness, constrained by well-defined rules, can produce works that appear deliberate and creative.

While generative art has seen various implementations, existing tools are often incomplete, scattered, or too abstract, making them inaccessible for artists and developers to experiment with easily.
The main goal of our work is to offer a simple yet powerful Python-based tool that enables users to create random yet aesthetically pleasing art.
\Samila, derived from the Farsi word meaning ``the one who put SURMI,'' embodies this philosophy by providing an intuitive interface for generating artistic figures through consistent random perturbations.
\Samila, our application for generative art, builds on this foundational ethos by mapping a two-dimensional Cartesian space to an arbitrary two-dimensional space using two mathematical pseudo-random functions.
Within Dorin’s framework, \Samila’s operation can be described as follows:
Each point (entity) in the Cartesian space is randomly displaced to a new coordinate (process) through a mathematical function controlled by random seeds (environmental interactions).
This process results in a visually compelling 2D figure (sensory outcome).
An example of generated \Samila art is in Figure \ref{fig:example_fig}.
The generation of Samila arts relies on controlled randomness, where the use of random seeds ensures that each output is both unique and non-reproducible without access to the specific seeds used.
This balance of unpredictability and control highlights the essence of generative art, as it marries algorithmic precision with creative spontaneity.

We demonstrated that \Samila’s generated output is uniquely determined by two random generation seeds, making regeneration nearly impossible without both due to the high degree of randomness.
Additionally, we observed that preserving the seed for the random generation function while altering the point generation functions produces distinct artworks that share common graphical characteristics, forming a cohesive visual family.

In the following sections, we first review related work (Section \ref{sec:rw}), providing context for our approach. Next, we delve into the structure of \Samila\footnote{Version 1.5: \url{https://zenodo.org/records/14721856}}, detailing the artwork generation process, the role of randomness, and the mechanisms ensuring reproducibility. We also highlight additional features that enhance usability. This section concludes with a step-by-step example, demonstrating the complete process of generating \Samila art from scratch (Section \ref{sec:program-struct}).
Finally, we explore potential future directions that build upon our work (Section \ref{sec:discuss}) and discuss its broader applications and impact on the community (Section \ref{sec:app}).

%% file: sections/rw.tex
\section{Related Work}
\label{sec:rw}
Generative art spans from classical randomness-based methods to modern AI-driven frameworks.
Early experiments in chance and automation, from Dada to Abstract Expressionism, laid the groundwork for computational approaches.
Today, algorithms and AI extend these traditions and blend randomness with structure.
In this section we trace the generative art’s evolution, from physical gestures to digital synthesis.

Artistic practitioners have experimented with generative techniques long before the digital age. Some scholars trace chance-based art back to Dada collages, while others emphasize surrealist automatism as a precursor to present-day procedural creativity. Abstract Expressionists such as Arnulf Rainer, Jackson Pollock, Mark Rothko, and Alexander Cozens used action painting to integrate spontaneous, near-random paint movements, revealing how gestures combined with natural forces like gravity could establish a partially autonomous system.

Jackson Pollock, a central figure in Abstract Expressionism, developed his signature drip technique, wherein paint was flung or dripped onto the canvas, allowing gravity and the viscosity of paint to influence the final outcome~\cite{rampley1996identity,schubert2013role}. His method underscored the importance of controlled randomness, a core tenet of generative aesthetics. Similarly, Arnulf Rainer embraced chance and automation through his over-painting process, layering chaotic, expressive brushstrokes over existing images—sometimes with his eyes closed—to deliberately reduce conscious control~\cite{schonlieb2013random}. His work parallels contemporary algorithmic art, where randomness is guided by structural constraints.

Mark Rothko took a different approach to generative composition, emphasizing large fields of color that interacted through subtle transitions and layering. Although not as explicitly process-driven as Pollock or Rainer, Rothko's color field painting explored emergent form through the optical blending of hues and edge diffusion, much like how digital generative art employs gradients and overlays to create complex, layered compositions~\cite{charlin2020price,wilkin2007color,adkins2008towards}. Meanwhile, Alexander Cozens, an 18th-century artist and theorist, devised a method of “blot drawing,” where abstract ink blots served as a starting point for landscapes, encouraging artists to find forms within chaos~\cite{cramer1997alexander,kuntz1966art}. This technique, reminiscent of modern procedural generation, foreshadowed how algorithmic randomness can serve as a foundation for structured aesthetic outcomes.

Opposed to contemporary AI-based generative art applications, traditional generative art often relies on computational algorithms that incorporate randomness as a core principle. Researchers have examined the mathematical underpinnings of randomness in visual arts, contrasting natural randomness with computational designs and tracing the evolution of randomness-based paintings~\cite{schonlieb2013random}.

Recent advancements in artificial intelligence have further expanded the possibilities for generative art. Researchers have explored using Python as a bridge between conventional art, design, and generative techniques, adopting practice-based methodologies to demonstrate fundamental coding principles for art creation~\cite{soikun2023understanding,babcock2021generative}. AI-driven frameworks, such as neural style transfer and Generative Adversarial Networks (GANs), have introduced new forms of artistic expression, where models learn to synthesize images that reimagine existing styles.

Parallel to these developments, several creative coding frameworks have emerged to facilitate generative art production. OpenFrameworks\footnote{https://openframeworks.cc/}, Cinder\footnote{https://libcinder.org/}, and Processing\footnote{https://processing.org/} provide robust libraries that enable both beginners and experienced artists to experiment with computational aesthetics.
These platforms increase innovation in generative techniques, offering tools for designing unique, algorithmically driven artworks.

One closely related project is the \emph{generativeart}\footnote{https://github.com/cutterkom/generativeart} package for R, which generates random abstract art pieces. However, its functionality is limited to R which is a hard programming language for users especially artists to interact with.
To address this, we developed \samila---a Python-based alternative that expands accessibility and customization options.
By incorporating principles of randomness and structured unpredictability, \samila builds on this rich artistic backbone, offering both a practical toolkit for computational exploration and a philosophical lens into the enduring value of chance and rules in art-making.

Early and contemporary explorations highlight the persistent appeal of complexity, spontaneity, and emergent form~\cite{slifkin2011tragic}. From Pollock’s paint-splattered canvases to modern AI-generated compositions, generative art consistently reflects an underlying philosophy: the creative process can be as integral as the final work. In each paradigm, a degree of autonomy is handed over to non-human forces—whether physical phenomena, random number generators, or machine-learning models. This approach challenges conventional notions of authorship and control, positioning the artist as a facilitator or orchestrator of emergent forms. Through \samila, we aim to contribute to this evolving discourse, providing an intuitive, open-ended system for both artistic exploration and computational creativity.

%% file: sections/program-struct.tex
\section{Program Structure}
\label{sec:program-struct}
\samila’s core architecture revolves around the interplay of mathematical functions and randomness to create generative art.
It uses randomized parameters, which dictate the behavior of the mathematical functions used to generate points.
Each execution with different random seeds results in unique, non-replicable artworks.
Thousands of points, arranged in order within a space, would transform into another space using $f_1(x, y)$ and $f_2(x, y)$, forming intricate patterns and designs.
To enhance the creative process, \samila offers extensive customization options, including the ability to define color schemes, adjust dot sizes, toggle grids, and refine the overall aesthetic.
Additionally, \samila supports exporting the generated artwork in various image formats and allows users to save the underlying mathematical parameters, ensuring reproducibility and enabling collaboration. An example of generating a random \samila artwork is shown below. It's as simple as three lines: initializing, generating, and plotting.

\begin{lstlisting}[language=Python]
from samila import GenerativeImage

g = GenerativeImage()
g.generate()
g.plot()
\end{lstlisting}

\subsection{Process}
The process of generating \samila artwork starts with initializing some steering parameters.
The artwork will then be generated given those parameters and plotted in a custom format.
A \samila artwork can be identically defined as a tuple $S$ as follows.
\begin{equation}
    \label{eq:samila}
    S = (A, C, c_{b}, p_{s}, p_{m}, p_{t})
\end{equation}
Where $A = \{(x_i, y_i)~|~ 1 \leq i \leq N\}$ is a list of points with size $N$ including first coordinate $x_i$, second coordinate $y_i$, and $C=\{c_i~|~ 1 \leq i \leq N\}$ is the set of corresponding colors for points, background-color $c_{b}$, point sizes $p_{s}$, point marker $p_{m}$, and point thickness $p_{t}$.
In following paragraphs we explain the process of generating instances of \samila generative artworks step-by-step.

\textbf{Parameter Initialization.}
In this first step, we require functions $f_1: R^2 \rightarrow R $ and $f_2: R^2 \rightarrow R$ that will be further used for random point generation.
Points will be mapped from the initial space to the later space using those two functions, i.e, points coordination in the later space will be determined by these two functions.
Additionally, to give the user the option to generate artwork randomly without any interventions, the user can let \samila generate random functions as well.
We detail this process in Section \ref{sec:random-function-gen}.

\textbf{Generation.}
Generating a \samila artwork starts with a set of $N$ points in two dimensional prime space, $A_0 = \{(x_i, y_i)~|~1 \leq i \leq N\}$.
Each point will be projected into another space, called the latter space $A_1$, using two functions, $f_1$ and $f_2$, which can be both deterministic or random.
We call this process of mapping between $A_0$ and $A_1$ ``generation'' because the main process of generating the artwork happens here.
In \samila we use a square of $A_0 = I \times I$ as our starting point set, where $\times$ is Cartesian production~\cite{dwyer2016systems} of between two sets and $I$ is defined as below.
\begin{equation}
    I = \{\verb|start| + k\times\verb|step|~|~0 \leq k < \lfloor\frac{stop - start}{step}\rfloor\}
\end{equation}

\begin{figure}
    \centering
    \includegraphics[width=\linewidth]{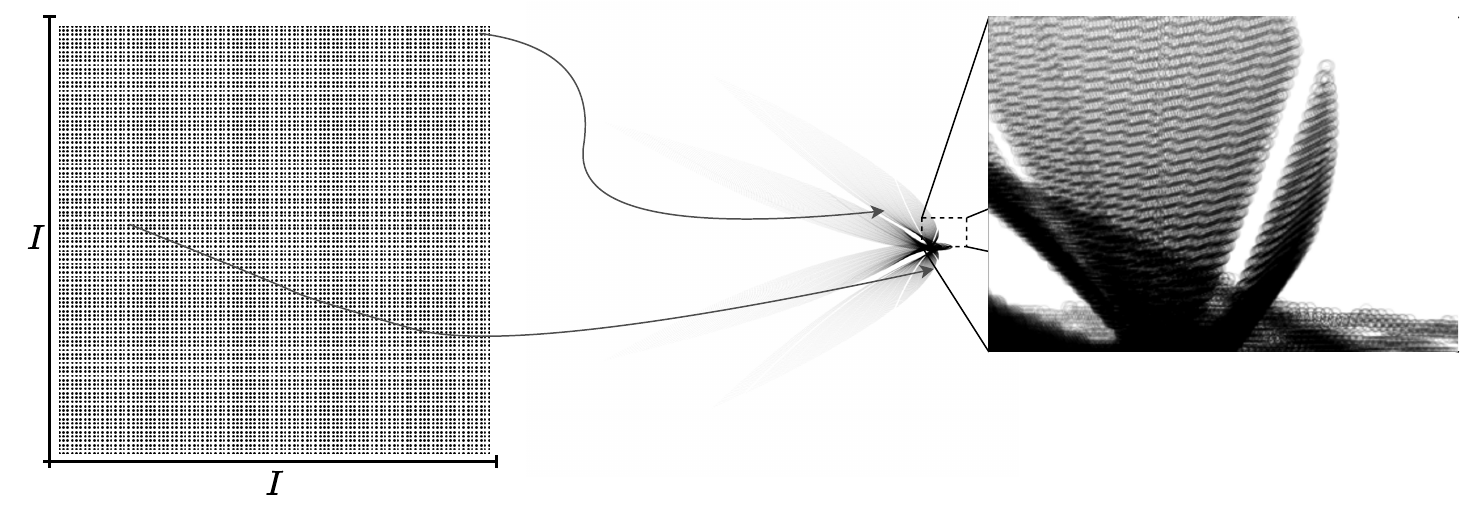}
    \caption{\samila artwork generation process. It started from a square of dense points and transformed them into the latent space to generate the artwork.}
    \label{fig:samila-transformation}
\end{figure}

In this part user can steer the generation with multiple parameters: \verb|seed| is the random seed that is used during the generation process, \verb|start|, \verb|step|, \verb|stop| are parameters that determines the size and granularity of the initial set of points, and \verb|mode| which is the parameter that determines the way that $A_1$'s point coordination is being determined from $A_0$ given $f_1$ and $f_2$. In the table below we detailed this relation.
\input{tables/generation-mode}

\begin{figure}
    \centering
    \includegraphics[width=\linewidth]{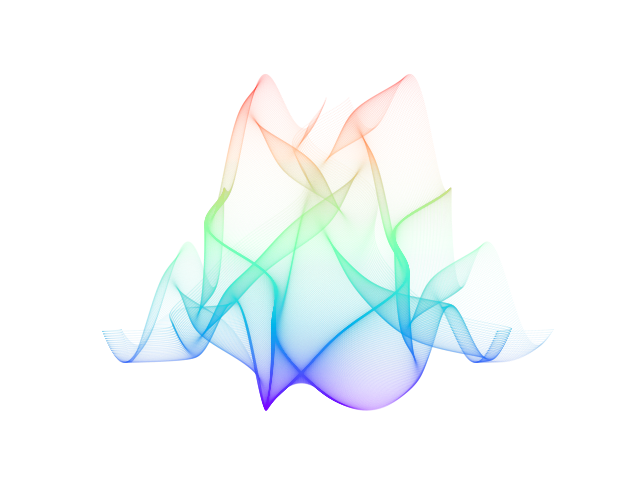}
    \caption{An example of \samila artwork which has different point colors. This example was generated by user ``meidefr'' in our Discord channel.}
    \label{fig:color-range}
\end{figure}

\textbf{Plotting.}
The last step in making the \samila artwork is to plot it.
The generated point set, $A_1$, can be presented in multiple different ways.
The user can control the plotting process by \verb|color| and \verb|cmap| parameters.
\samila supports both constant colors for all points, $c_i = c$, and a list of colors.
That list of colors can also be defined as a function of point coordination so one have $c_i = f_c(x_i, y_i)$ therefore giving the opportunity for more explorations.
An example of an artwork that has different point colors based on the vertical coordinate is provided in Figure \ref{fig:color-range}.

\begin{figure*}
    \centering
    \includegraphics[width=\linewidth]{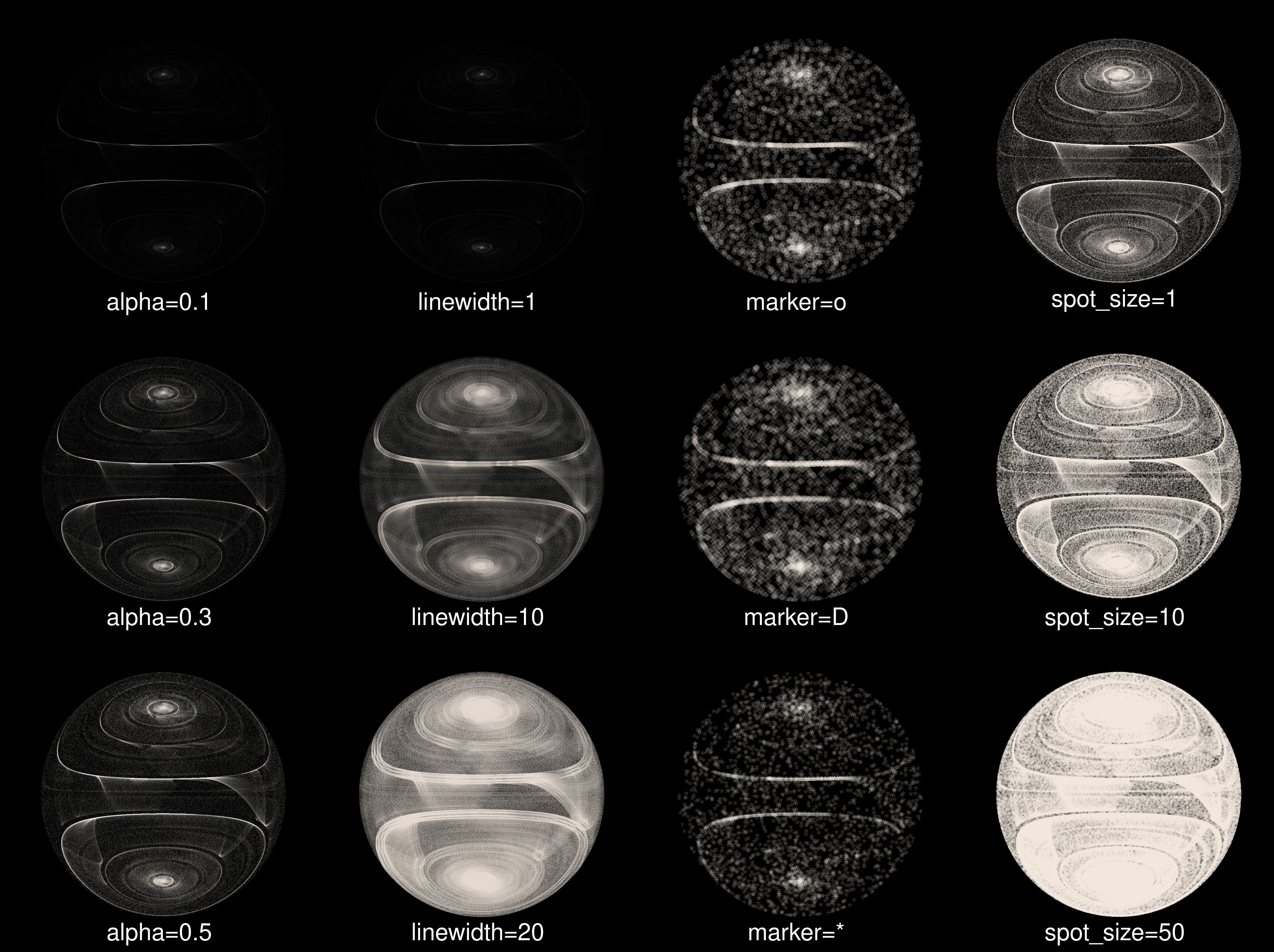}
    \caption{Effect of different plotting parameters on the artwork. In each of the columns, all the parameters except one that is mentioned are changed in different rows. In the column for the marker, we increased spot size and decreased step size to have fewer points with a bigger size for the sake of presenting the effect.}
    \label{fig:paramter-range}
\end{figure*}

In addition to color users can control background color \verb|bgcolor|, and point features such as size \verb|spot_size|, marker \verb|marker|, boldness \verb|linewidth|, and transparency \verb|alpha|.
An overview of these parameters can be seen in Figure \ref{fig:paramter-range}.

Finally, when the building blocks of the artwork are set, users can use different projections of points, these projections are \verb|matplotlib|'s projections and results in different artworks.
This can be controlled using \verb|projection| parameter in \verb|plot| function, which accepts projection enumerations.
For example when \verb|polar| projection sets $(x,y)=(\theta,r)$ and plot $(y\cos x,y\sin x)$ and \verb|lambert| projects sets $(x,y)=(\sin\theta,y)$ and plot $(\arcsin x, y)$.  
In addition to those predefined projection users can rotate the generated artwork by \verb|rotation| parameter.

\subsection{Randomness and Reproducibility}
\label{sec:randomness-repr}
Randomness and reproducibility are central to \samila artwork generation.
A specific artwork can be recreated only with its key (configuration file or seed).
The process is random because the configuration cannot be inferred from the artwork, and reproducible because the same configuration guarantees identical output.
In essence, it establishes a near one-way mapping between configurations and artworks.
In this section, we describe the two processes—random function generation and point generation—that enable us to ensure both properties.

\textbf{Random Function Generation.}
\label{sec:random-function-gen}
We proposed a novel method for generating random functions.
Although \samila gives you the option to change the hyperparameters of the random function generation, we experimentally tweaked them so that the generated artworks are ``beautiful.''
We noticed that such an artwork happens when the randomness is controlled, i.e, exists but not too much which resulting into chaos. This aligns with Galanter's findings regarding the type of randomness contained in a generative art~\cite{galanter2016generative}.

Before starting with the random function generation process we would like to present some preliminary notations.
A pseudo-random function $f_r$ is a function that would generate a random number between a specified range given a seed.
An example of such a function is a Gaussian pseudo-random number generator with $\mu=0$ and $\sigma=1$ that generates a sample of a Gaussian random variable with mean zero and variance one. We call a finite set of such functions a pseudo-random function family $\mathcal{F}_r$.
Similarly deterministic functions $f: \mathbf{R} \rightarrow \mathbf{R}$ can form a family of functions as a finite set $\mathcal{F}$.
The set of arguments $\mathcal{A}$ is the finite set of functions $f: \mathbf{R}^2 \rightarrow \mathbf{R}$. For example, $x \times y \in \mathcal{A}$ determines how $x$ and $y$ are put into functions' arguments.
The operator set $\mathcal{O}$ is a finite set of two-operand associative operators such as $+$ and $/$.

We generate random mathematical equations as strings that include terms $t_i$ connected by operators $o_i$.
First, a random integer $n$ is uniformly selected from $n \in_U [C_{min}, C_{max}]$ to determine the number of terms in the equation, which is a controller for the complexity of the equation.
Then, as a source of randomness inside function generations, we chose a pseudo-random function $f^r$ driven from a family of base pseudo-random functions $\mathcal{F}^r$.
For each term $t_i (1 \leq i \leq n)$, a random integer $d_i$ is uniformly selected from $d \in_U [D_{min}, D_{max}]$ which determines the number of recursions inside this term, i.e, the depth of random function generations in arguments of $t_i$.
For each recursion, a function $f$ will be randomly sampled from a family of functions deterministically defined functions $\mathcal{F}$.
Then the selected pseudo-random function $f^r$ will be multiplied in the $f$.
This process repeated $d$ times and the final input argument $a_i$ will be randomly picked from $\mathcal{A}$ until we have $t_i = f_rf_i^1(f_rf_i^2(\dots f_rf_i^d(a_i) \dots))$, where $f_r$ is the random function that has been chose at the beginning of generation.
This term then appends to the final string with an operator $o_i$ uniformly selected from a set of operators $\mathcal{O}$.
Yielding to a string which has a structure like $s = t_1~o_1~t_2~o_2~\dots~o_{n-1}i_n$, with a probability of $p$ another $f$ would be sampled from $\mathcal{F}$ and the results would be the multiplication of $f_r$ into that function using $s$ as argument $f_rf(s)$.
Below we have a pseudo-code version of the random equation generator (Algorithm \ref{alg:random-eq-gen}).

\begin{algorithm}
\caption{Random Equation Generation}
\label{alg:random-eq-gen}
\begin{algorithmic}[1]
\Require Constants $C_{min}, C_{max}, D_{min}, D_{max}$
\Require Function families $\mathcal{F}_r$ (pseudo-random), $\mathcal{F}$ (deterministic)
\Require Argument set $\mathcal{A}$, Operator set $\mathcal{O}$
\Ensure Randomly generated mathematical equation as a string

\State Initialize result string $S \gets ``"$
\State Select a random integer $n \sim U(C_{min}, C_{max})$
\State Select a pseudo-random function $f_r$ from $\mathcal{F}_r$
\For{$i = 1$ to $n$}
    \State Select a random integer $d \sim U(D_{min}, D_{max})$
    \State Select a random argument $a \in \mathcal{A}$
    \For{$j = 1$ to $d$}
        \State Select a function $f$ from $\mathcal{F}$
        \State Apply function composition: $a \gets f_r \cdot f(a)$
    \EndFor
    \State  $t_i \gets a$
    \State $S \gets S + t_i$
    \If{$i < n$}
        \State $S \gets S + o\in \mathcal{O}$
    \EndIf
\EndFor
\If{with probability $p$}
    \State Select a function $f$ from $\mathcal{F}$
    \State $S \gets f_r \cdot f(S)$
\EndIf
\State \Return $S$
\end{algorithmic}
\end{algorithm}

Relaxing the constant parameters for the numbers of terms $C_{min}, C_{max}$ and depth $D_{min}, D_{max}$, we can then describe the family of generated functions of this algorithm by a context-free grammar like below.
$S, F_r, F, S', T, O, A$ are all non-terminal variables and $f_r^i, f^i, o^i, a^i$ are terminal variables.
Meanwhile, $|.|$ indicated the size of a set, and $(.)$ indicates the argument of the function.
Additionall,y a pseudo-random function $f_r$ is sampled from $\mathcal{F}_r$, which is constant through the generation and can be considered a terminal variable.

\begin{align*}
S &\to f_rF(S') \OR S'\\
S' &\to T \OR TOT \\
T &\to A \OR f_rF(T)\\
F &\to f^1\OR \cdots \OR f^{|\mathcal{F}|} \\
O &\to o^1 \OR \cdots \OR o^{|\mathcal{O}|}\\
A &\to a^1 \OR \cdots \OR a^{|\mathcal{A}|}\\
\end{align*}

While users can choose their parameters for the generation, \samila provides a set of pre-defined parameters.
We tweaked these parameters until we reach a point that we were getting more ``interesting'' results. However, since ``interesting'' does not have a specific meaning, these parameters are optimized subjectively based on this paper's writers' perspective and are not the global-optimal parameters --- if there exist any.
The parameters that we used for each of the sets are presented in Table \ref{tab:math_sets}.
The definitions for the following random functions are presented by Thomopoulos et. al~\cite{thomopoulos2018probability}.
The seed that is used for random function generation is controlled by \verb|func_seed| or will be generated randomly if it's not given.
\input{tables/math_set}





We set the minimum depth of each term to $D_{min}=1$ and the maximum depth to $D_{max}=2$. Also, the minimum number of terms is $C_{min}=1$ and the maximum to $C_{max}=|F|+1=14$.

An example of a randomly generated function by default parameters in \samila can be as follows where $f_r = Uniform(-1,1)$.
\begin{align*}
    f = f_r \lceil y \rceil - f_r y^2 + f_r |y - x|
\end{align*}

\begin{figure}[h]
    \centering
    \begin{forest}
        [S
            [S'
                [T
                    [$f_r$]
                    [F
                        [$\lceil$]
                        [T
                            [A [$y$]]
                        ]
                        [$\rceil$]
                    ]
                ]
                [O [$-$]]
                [T
                    [T
                        [$f_r$]
                        [F
                            [(]
                            [T
                                [A [$y$]]
                            ]
                            [)$^2$]
                        ]
                    ]
                    [O [$+$]]
                    [T
                        [$f_r$]
                        [F
                            [$|$]
                            [T
                                [A [$y-x$]]
                            ]
                            [$|$]
                        ]
                    ]
                ]
            ]
        ]
    \end{forest}
    \caption{Parse tree representation of the equation}
    \label{fig:parse_tree}
\end{figure}
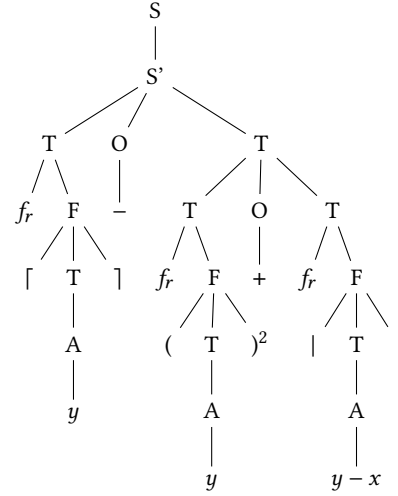

A parsing tree of this function is also proposed in Figure \ref{fig:parse_tree}. Note that the parsing three is not unique so there could be other parsing threes for this generated function.

\textbf{Random Point Generation.}
\label{sec:random-point-gen}
The process of generating random points from given functions is simpler.
Here we assume that two projecting functions $f_1$ and $f_2$, in addition to a starting set of points $A_0$ are given.
Now given a random seed that is either inputted by the user \verb|seed| or generated randomly, \samila would transfer the point from $A_0$ to $A$.

These two control parameters --- \verb|func_seed| and \verb|seed| --- not only deterministically determine the generated artwork, but also act as a key so that generating the same artwork is impossible not having either of them.
Furthermore, these seeds can be any Python object and they are not only numbers.

\begin{figure*}
    \centering
    \includegraphics[width=\linewidth]{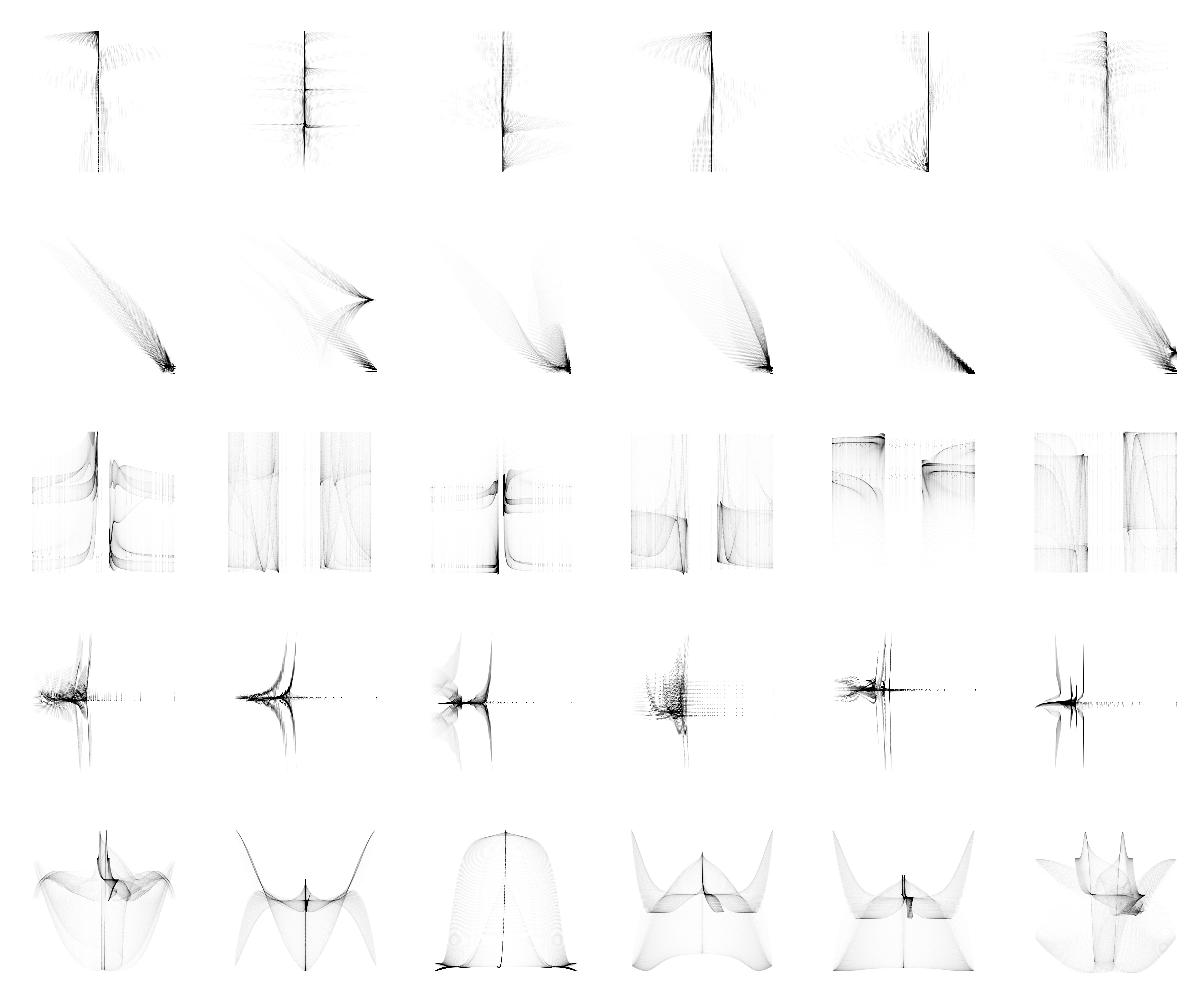}
    \caption{Effect of varying function seed (rows) and seed (columns) of \samila artwork. Fixing a function seed would generate a family of artworks which share specific graphical features. Function seeds used for the rows are [41868, 20523, 30891, 44863, 5682] and the seed used for each column is [10798, 33914, 39080, 68261, 76731, 90039, 94846]}
    \label{fig:funcseed-seed}
\end{figure*}

We observed that fixing the \verb|func_seed|, i.e., fixing the projection functions, and changing the \verb|seed| would change the generated artwork in a way that they all represent a family of artworks rather than completely different artworks.
This notion of continuity can be observed from Figure \ref{fig:funcseed-seed}, where in one dimension we alter \verb|func_seed| and in another we're changing \verb|seed|.

\subsection{Command-Line Interface (CLI)}
\input{tables/cli-example-params}
For easier access through command line tools like \textit{terminal} or \textit{cmd} we developed a command line interface.
With that, users can generate \samila artwork even without entering into an editor or using a Python interpreter. 
This will help them integrate \samila into their workflow more easily by calling its CLI command through the operating system.
An example of this interface is shown as below.
\begin{lstlisting}[language=Bash]
samila --verbose --no-display --color=red \
       --bgcolor=black --rotation=30 --projection=polar \
       --mode f2_vs_f1 --save-image test.png
\end{lstlisting}

A summary of what each parameter does is presented in Table \ref{tab:cli-arams}.
The final artwork would finally be rotated by $30^{\circ}$ and saved in \verb|test.png|.

\begin{figure*}[h]
    \centering
    \begin{minipage}{0.32\textwidth}
        \centering
        \includegraphics[width=\linewidth]{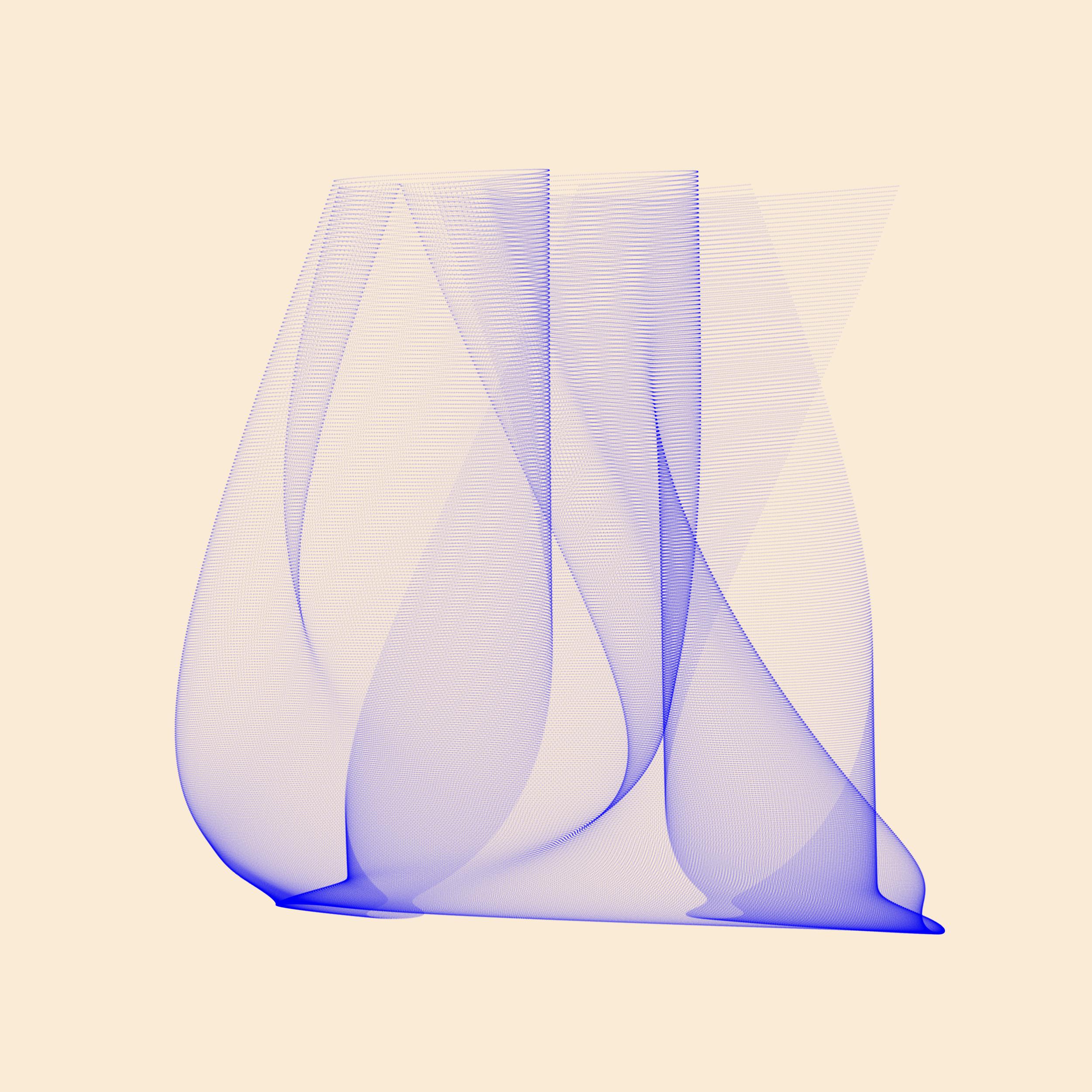}

    \end{minipage}
    \hfill
    \begin{minipage}{0.32\textwidth}
        \centering
        \includegraphics[width=\linewidth]{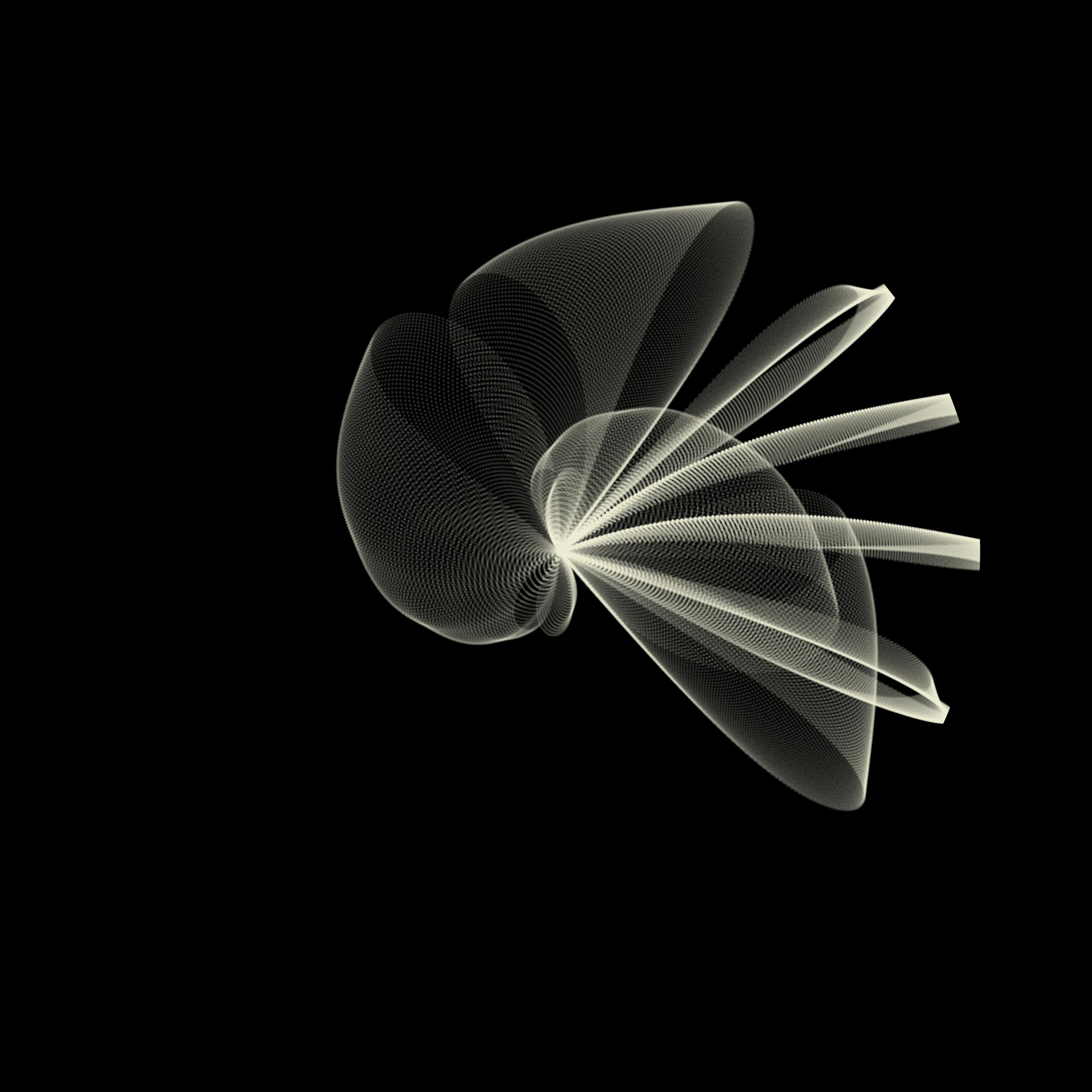}
    \end{minipage}
    \hfill
    \begin{minipage}{0.32\textwidth}
        \centering
        \includegraphics[width=\linewidth]{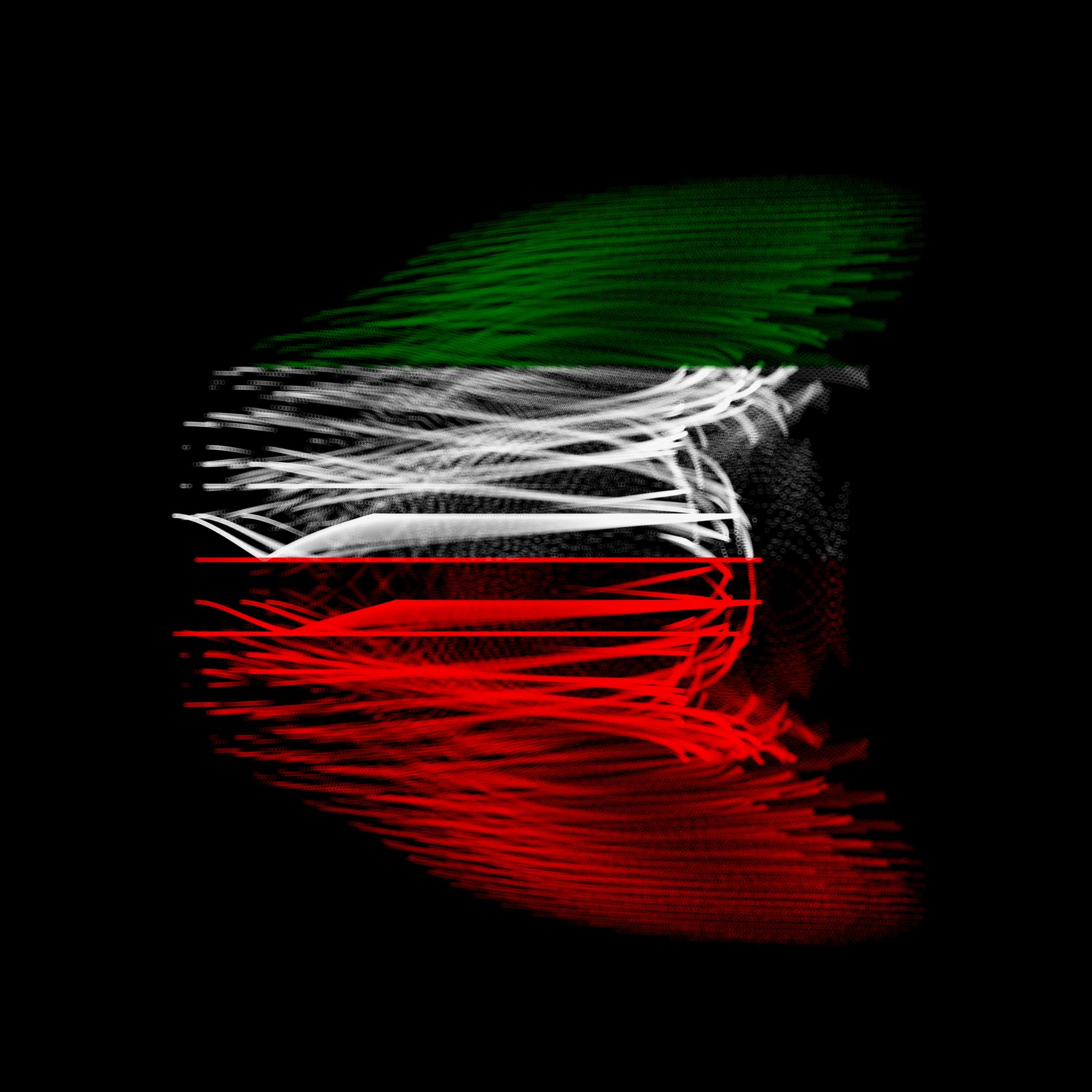}
    \end{minipage}
    \caption{Examples of \samila generative art with their configuration file for regenerability. Point colors $C$ are cut down to two elements for easier representation. Configuration files are presented in Appendix\ref{sec:app:conf}.}
    \label{fig:samila-examples}
\end{figure*}

\subsection{I/O}
In section \ref{sec:randomness-repr} we discussed that the process of random \samila artwork generation is controlled by two parameter --- one random seed for the function and other for point generation.
However, to make the reproducibility of \samila arts more transparent and accessible, we defined three input/outputting methods with differnet levels of control over reconstruction.
From Equation \ref{eq:samila}, we remember that a \samila artwork $S$ has $A, C, c_{b}, p_{s}, p_{m}, p_{t}$ as components. The level of control the user would have over the reservation of the artwork depends on how processed the saved information.

\textbf{Image File.}
Users can easily save images in formats like \verb|png|, \verb|jpg|, \verb|svg|, \verb|pdf|, and et. After saving \samila artwork $S$, other users have the least accessibility to regenerate or edit the artwork due to the complex underlying structure.

\textbf{Data File.}
Users can also save the projected set of points $A$ as data.
This way, they can plot the set of points into different spaces with different rotations and different coloring, and formatting.
For example having the data file of an artwork like Figure \ref{fig:color-range}, you can rotate it by $90^{\circ}$ and color it all to black.
Data file falls in the middle regarding the option of regenerability.

\textbf{Configuration File.}
The most powerful file which lets user to regenerate the exact same \samila artwork elsewhere, is the configuration file. This file includes all the information, such as random seeds, the initial set $A_0$, transformation functions, and plotting settings.
Given this file one can regenerate the exact same artwork as it's generated with.
Therefore, configuration files act as a unique signature of \samila's artwork and have an equivalence with $S$.
Some examples of the configuration file's general structure is provided in Appendix\ref{sec:app:conf}. You can regenerate the same images using those files.

\subsection{Examples}
\samila's outputs can sometimes be eye-catching. In Figure \ref{fig:samila-examples} you can see three different randomly generated random artworks with their configuration file.
The left-most image is in \verb|rectilinear| projection with blue point colors on a  \verb|antiquewhite| background.
The middle one is in \verb|polar| projection with beige point colors and black background.
And the right-most one, which is again in rectilinear projection, has different coloring based on $y$ coordination of points starting from red changing into white and ending in green with black background.

%% file: tables/generation-mode.tex
\begin{table}[tbh]
\caption{Mapping modes to generated data points $(x_i', y_i') \in A_1$ from $(x_i, y_i) \in A_0$ and given $f_1$ and $f_2$ functions.}
    \centering
    \begin{tabular}{|c|c|}
    \hline
    \textbf{mode} & \textbf{Formula for transformation} \\
    \hline
     \verb|F1_VS_F2|    &  $(f_1(x_i, y_i), f_2(x_i, y_i))$ \\
     \verb|F2_VS_F1|    &  $(f_2(x_i, y_i), f_1(x_i, y_i))$ \\
     \verb|F2_VS_INDEX| &  $(f_2(x_i, y_i), i)$ \\
     \verb|F1_VS_INDEX| &  $(f_1(x_i, y_i), i)$ \\
     \verb|INDEX_VS_F1| &  $(i, f_1(x_i, y_i))$ \\
     \verb|INDEX_VS_F2| &  $(i, f_2(x_i, y_i))$ \\
     \verb|F1_VS_X1|    &  $(f_1(x_i, y_i), x_i)$ \\
     \verb|F2_VS_X1|    &  $(f_2(x_i, y_i), x_i)$ \\
     \verb|F1_VS_X2|    &  $(f_1(x_i, y_i), y_i)$ \\
     \verb|F2_VS_X2|    &  $(f_2(x_i, y_i), y_i)$ \\
     \verb|X1_VS_F1|    &  $(x_i, f_1(x_i, y_i))$ \\
     \verb|X1_VS_F2|    &  $(x_i, f_2(x_i, y_i))$ \\
     \verb|X2_VS_F1|    &  $(y_i, f_1(x_i, y_i))$ \\
     \verb|X2_VS_F2|    &  $(y_i, f_2(x_i, y_i))$ \\
     \hline
    \end{tabular}
    \label{tab:mode_mappings}
\end{table}

%% file: tables/math_set.tex


\begin{table*}[h]
    \centering
    \caption{Mathematical Sets and Their Elements in \Samila}
    \renewcommand{\arraystretch}{1.1}
    \resizebox{\textwidth}{!}{
    \begin{tabular}{|c|c|c|c|c|c|c|c|c|c|c|c|c|c|c|c|c|c|c|c|c|c|}
        \hline
        \textbf{Set} & \multicolumn{21}{c|}{\textbf{Elements}} \\
        \hline
        $\mathcal{A}$ &
            $x \times y$ &
            $x$ & 
            $y$ &
            $\frac{1}{x}$ &
            $\frac{1}{y}$ &
            $\frac{x}{y}$ &
            $y - x$ &
            $x - y$ &
            $x + y$ &
            $x^3$ &
            $y^3$ &
            $x^2$ &
            $y^2$&
            $x^2 \times y$ &
            $y^2 \times x$ &
            $x^2 + y^2$ &
            $y^2 - x^2$ &
            $x^2 \times y^3$ &
            $x^3 \times y^2$ &
            $x \times y^3$ &
            $y \times x^3$ \\
        \hline
        $\mathcal{F}$ &
            $\tanh(x)$ &
            $\cos(x)$ &
            $\sin(x)$ &
            $x$ &
            $|x|$ &
            \multicolumn{2}{c|}{$\lceil x \rceil$} &
            \multicolumn{2}{c|}{$\lfloor x \rfloor$} & 
            \multicolumn{2}{c|}{$\tan(x)$} &
            \multicolumn{2}{c|}{$\erf(x)$} &
            \multicolumn{2}{c|}{$\sqrt{|x|}$} &
            \multicolumn{2}{c|}{$\log(|x|+1)$} &
            \multicolumn{2}{c|}{$\arccosh(|x|+1)$} &
            \multicolumn{2}{c|}{$\arcsinh(x)$} \\
        \hline
        $\mathcal{F}_r$ &
            \multicolumn{4}{c|}{Uniform($a=-1, b=1$)}  &
            \multicolumn{4}{c|}{Gaussian($\mu=0, \sigma=1$)} &
            \multicolumn{4}{c|}{Betavariate($\alpha=1, \beta=1$)} &
            \multicolumn{4}{c|}{Gammavariate($\alpha=1, \beta=1$)} &
            \multicolumn{5}{c|}{Lognormvariate($\mu=0, \sigma=1$)} \\
        \hline
        $\mathcal{O}$ &
            \multicolumn{5}{c|}{$+$} &
            \multicolumn{5}{c|}{$-$} &
            \multicolumn{5}{c|}{$\times$} &
            \multicolumn{6}{c|}{$/$} \\
        \hline
    \end{tabular}
    }
    \label{tab:math_sets}
\end{table*}

%% file: tables/cli-example-params.tex
\begin{table}[h]
    \centering
    \caption{Summary of Parameters}
    \renewcommand{\arraystretch}{1.1}
    \resizebox{\linewidth}{!}{
    \begin{tabular}{|l|l|}
        \hline
        \textbf{Parameter} & \textbf{Description} \\
        \hline
        \texttt{verbose} & Logs all generation information \\
        \hline
        \texttt{no-display} & Prevents the generated image from being displayed \\
        \hline
        \texttt{bgcolor=black} & Sets the background color to black \\
        \hline
        \texttt{color=red} & Sets the dot color to red \\
        \hline
        \texttt{mode f2\_vs\_f1} & Sets the projection mode to polar with $f_2$ and $f_1$ as controllers \\
        \hline
    \end{tabular}}
    \label{tab:cli-arams}
\end{table}

%% file: sections/discussion.tex
\section{Discussion}
\label{sec:discuss}

Generative art, whether derived from paint drips or Python code, thrives on a delicate balance between structure and spontaneity.
The tension between the artist’s intentions and the system’s inherent unpredictability consistently raises key questions regarding authorship, control, and creativity. 

\textbf{Authorship.}
From an artistic perspective, such practice challenges viewers and creators alike to appreciate that the ``emergent'' quality of the artwork is no longer the sole product of an individual’s hand but the result of interacting rules, forces, or computations that operate autonomously once set in motion. In line with Galanter’s observations~\cite{galanter2016generative}, the generative process shifts the attention from the final static outputs to the dynamic, evolving procedures, stimulating an ongoing inquiry into how much control an artist should relinquish versus how much unexpected novelty arises when rules interact with chance.

\textbf{Creativity.}
Within this context, one might also question the aesthetic criteria by which these works are judged. Unlike traditional painting or sculpture, a generative piece can produce innumerable permutations, each one valid yet fleetingly unique. The process-oriented emphasis seen in Pollock’s ``action painting'' resurfaces in digital generative systems, where execution and iteration sometimes supersede the significance of any single outcome. As a result, the artwork’s conceptual weight can reside in the algorithm or physical process itself, prompting a reexamination of what constitutes ``the work'' in a world of endless computational variation~\cite{dorin2013chance,meng2023predicting}.

\textbf{Provocation.}
Generative art also presents a unique opportunity for visual provocation and engagement.
The unpredictable emergence of form, texture, and pattern can make these works visually compelling, as they resist immediate categorization and encourage prolonged observation~\cite{meng2023predicting}.
In digital generative systems, variations in density, contrast, and randomness create an interplay between order and chaos, drawing the viewer into an active, dynamic visual experience.

\textbf{Chromatics.}
Another defining characteristic of generative art is its transition from monochromatic simplicity to complex, multi-colored compositions that generate spatial illusions~\cite{mai2010compositional}. Much like Mark Rothko’s layered color fields, digital generative pieces can manipulate hue, saturation, and transparency to create a sense of depth and motion within a two-dimensional plane. By adjusting color relationships dynamically, artists can achieve an optical push-and-pull effect, making compositions appear to shift, vibrate, or recede into space~\cite{kim2013study}.

\textbf{Off-Center.}
One intriguing aspect of Samila's generative process is its tendency to favor central compositions in \verb|polar| projection mode, where generated forms often emerge from the center of the image plane.
However, an important question in composition arises: ``\textit{Can we shift the focal weight of the artwork towards the corners or edges instead of the center?}''
In traditional painting and photography, off-center composition often creates a sense of movement, asymmetry, and narrative tension.
While most generative systems, including \Samila, start by mapping points symmetrically within a defined plane, adjusting parameters such as random seed distributions, density weighting, or projection offsets could allow the emergence of compositions that naturally originate from non-central regions of the canvas.

By experimenting with different function mappings in Samila, it is possible to bias generative compositions toward the periphery, creating focal points in the corners or along the edges instead of the traditional central cluster.
This approach mirrors certain principles in traditional art, such as the rule of thirds, where compositions are deliberately shifted away from the center to create a more engaging visual balance. Exploring off-center compositions within generative systems challenges the conventional aesthetic of symmetrical randomness and opens new possibilities for dynamic spatial relationships in algorithmic design.

%% file: sections/future-works.tex
\section{Future Work}
\textbf{Using AI for Generating Artworks.}  
Future developments could integrate artificial intelligence models, especially generative models such as Generative Adversarial Networks (GANs), into the generative process.
GANs consist of a generator network, which generates images, and a discriminator network, which attempts to distinguish between real data and data produced by the generator~\cite{goodfellow2020generative}.
By training GANs on large datasets of existing artworks or Samila-generated images, one could refine randomness, optimize parameter selection, and generate compositions that adhere to learned stylistic patterns.
Furthermore, techniques such as StyleGAN~\cite{karras2019style} could allow fine-grained control over image attributes, allowing users to explore variations in form, texture, and composition with greater precision. 

\textbf{Hack the Generation Process.}  
Currently, Samila operates in a one-way fashion, generating images from mathematical functions $f_1$ and $f_2$ without an inverse mechanism.
However, reconstructing these functions from a given image can be an interesting and challenging research question.
Future work could explore the use of Kolmogorov-Arnold networks (KAN)~\cite{liu2024kan} to approximate inverse mappings, allowing users to generate mathematical representations of existing artworks.
This would enable a more interactive and bidirectional creative process, where users could iteratively refine their generative artworks based on desired visual characteristics.  

\textbf{Objectifying Beauty.}  
Defining a quantitative metric for aesthetic appeal is an open problem.
A promising approach involves collecting human-annotated rankings of Samila-generated images and using these rankings to train machine learning models.
Using the underlying \samila generative functions as latent feature representations, we could identify mathematical properties, such as curvature or symmetry, that contribute to perceived beauty.
Large-scale user studies could also help establish a robust dataset in which images are ranked on a scale (e.g., 1 to 10). This ranking data could then be used to define an empirical beauty metric, guiding the development of more visually compelling generative models.

%% file: sections/limit.tex
\section{Limitation}
One fundamental limitation of this study is the inherent subjectivity in evaluating the aesthetic quality of generative artworks.
Since beauty is a subjective concept, determining whether a generated piece is visually compelling remains an open challenge.
To mitigate this, we employed a negotiated agreement between two authors when making adjustments to the random generation process, ensuring a more balanced and deliberate refinement of outputs.

Additionally, our findings are not supported by rigorous statistical analysis.
While we formulated hypotheses about the visual characteristics of generated artworks, these were not quantitatively validated.
Future work will incorporate human feedback and empirical evaluation methods to strengthen the analytical foundation of our claims.

Finally, the current implementation of our system has certain software limitations.
For example, it does not yet support custom projections, which restricts the range of compositional structures that can be explored.
These and other technical constraints will be addressed in future iterations to enhance flexibility and user control over the generative process.

%% file: sections/application.tex
\section{Applications and Implications}
\label{sec:app}

Contemporary artists embrace process by experimenting with code, machine learning, or interactive installations.
Generative frameworks encourage collaborations with autonomous systems, enabling new modes of expression that blend the corporeal (physical paint, sculptural elements) with the virtual (algorithmic randomness, AI-driven transformations). 
Inline with that, \Samila offers visual artists an interactive platform to explore complex layering through controlled randomness.
Artists might begin with a simple monochromatic setup, gradually introducing layers of color or changing density and transparency settings to evoke a sense of space and movement within the composition.
The immediacy of visual feedback provided by \Samila accelerates visual thinking, provoking unexpected aesthetic insights through rapid experimentation.
Such a generative approach can act as a breakthrough for expressive artists seeking inspiration, where algorithmically produced outputs serve as starting points or even direct components in mixed-media projects.
Specifically, Samila offers artists an accessible gateway to generative exploration, allowing them to experiment intuitively with random seeds, mathematical functions, and projection methods.
By adjusting these parameters, artists can quickly iterate between compositions, discovering forms and structures that would be challenging to conceive manually.

\textbf{Visual Artists and Painters.}
Samila provides visual artists, particularly painters, with a powerful toolkit to augment traditional studio practices.
Painters can use generative outputs from Samila as preliminary sketches or compositional references, exploring dynamic, asymmetrical arrangements that depart from conventional centered formats.
For instance, by manipulating parameters, artists can quickly produce compositions that emphasize corner or edge-focused focal points, generating tension, movement, or visual intrigue uncommon in centrally oriented artworks.
Such explorations can serve as digital sketches or conceptual frameworks for physical paintings, allowing artists to experiment with asymmetrical balance, negative space, and rhythm before committing to canvas.
Additionally, Samila’s ability to transition smoothly from monochrome to complex, multi-layered color schemes offers painters an experimental ground for investigating optical illusions, spatial depth, and color relationships, echoing traditional practices such as glazing or layering in physical mediums.
Thus, Samila serves as a complement to human creativity, enabling artists to engage in a human-computer interaction framework that enhances artistic exploration through iterative visual experimentation.

\textbf{Education.}
\Samila presents an intuitive tool for educators aiming to demonstrate mathematical concepts, randomness, and algorithmic thinking in an interactive manner.
By visualizing mathematical functions and stochastic processes, students can gain a deeper understanding of topics such as probability distributions, function transformations, and computational geometry.
Furthermore, Samila's generative nature can be integrated into programming courses, allowing students to explore creative coding while reinforcing core programming concepts.

\textbf{Research.}
In computational creativity and generative design, Samila serves as a valuable tool for prototyping, visualization, and experimentation.
Researchers can use it to study emergent patterns, analyze the impact of different function compositions, or investigate the aesthetic properties of algorithmically generated forms.
Additionally, \Samila’s structured randomness provides a controlled setting for exploring algorithmic bias in generative art, offering insights into how different functions contribute to visual composition.

By bridging computational methods with artistic expression, Samila enhances interdisciplinary engagement, making generative art more accessible to a diverse audience.
Whether as an educational aid, a research instrument, or a creative catalyst, Samila continues to expand the possibilities of algorithm-driven artistic exploration.

%% file: sections/conclusion.tex

\section{Conclusion}
\Samila is a Python library for generative art that combines mathematical functions and randomness to create visually striking compositions.
It enables artists to experiment with structured randomness through parameters like random seeds and projection modes, allowing for interactive human-computer collaboration.
\Samila’s output is uniquely determined by two random seeds, making regeneration nearly impossible without both.
Additionally, by keeping one seed the same and adjusting point generation functions, distinct artworks with shared visual characteristics can be produced.
Beyond creative exploration, \Samila also serves as an educational tool for teaching mathematical and programming concepts, as well as a research platform for generative design.
Although it currently lacks custom projections and inverse function retrieval, future updates could incorporate AI-driven generation and aesthetic evaluation metrics, enhancing control and accessibility for users.

%% file: sections/appendix.tex
\onecolumn
\appendix
\section*{Appendix}

\subsection{Configuration Files}
\label{sec:app:conf}
In this section we provided the three configuration files we used for generating images in Figure \ref{fig:samila-examples}. You can load these files directly into \Samila package and make the same artworks.

\begin{lstlisting}[language=bash, caption={Configuration file for the left subfigure in Figure \ref{fig:samila-examples}.}]
{
    "matplotlib_version": "3.0.3",
    "f2": "random.betavariate(1,1)*math.cos(x+y)-random.betavariate(1,1)*math.log(abs(y**2)+1)+random.betavariate(1,1)*math.sin(x+y)+random.betavariate(1,1)*math.tanh(x**2)+random.betavariate(1,1)*math.sin(x+y)+random.betavariate(1,1)*math.erf((x**2)*(y**3))+random.betavariate(1,1)*math.erf(x)",
    "plot": {
        "alpha": 0.1,
        "linewidth": 0.04,
        "bgcolor": "antiquewhite",
        "color": "b",
        "projection": "rectilinear",
        "depth": 5,
        "spot_size": 0.77
    },
    "generate": {
    "f1": "random.gammavariate(1,1)*abs(y**2)-random.gammavariate(1,1)*math.sin(x)"
        "stop": 3.141592653589793,
        "step": 0.01,
        "seed": 778783,
        "start": -3.141592653589793
    },
}
\end{lstlisting}

\begin{lstlisting}[language=bash, caption={Configuration file for the middle subfigure in Figure \ref{fig:samila-examples}.}]
{
    "f2": "random.uniform(-1,1)*math.floor(x+y)-random.uniform(-1,1)*abs((x**2)*y)+random.uniform(-1,1)*math.sin(x*y)-random.uniform(-1,1)*math.cos((x**2)*(y**3))",
    "f1": "random.uniform(-1,1)*abs((x**2)*y)+random.uniform(-1,1)*math.cos(x-y)",
    "matplotlib_version": "3.0.3",
    "generate": {
        "step": 0.01,
        "stop": 3.141592653589793,
        "start": -3.141592653589793,
        "seed": 561872
    },
    "plot": {
        "color": "beige",
        "alpha": 0.1,
        "bgcolor": "black",
        "projection": "polar",
        "spot_size": 1
    }
}
\end{lstlisting}

\begin{lstlisting}[language=bash, caption={Configuration file for the right subfigure in Figure \ref{fig:samila-examples}.}]
{
    "f1": "random.gauss(0,1)*math.sqrt(abs(y*(x**3)))*random.gauss(0,1)*math.tanh((y**2)*x)/random.gauss(0,1)*math.sin(random.gauss(0,1)*math.atan(x**3))/random.gauss(0,1)*math.cos(y)+random.gauss(0,1)*math.asinh((y**2)-(x**2))+random.gauss(0,1)*math.cos(random.gauss(0,1)*math.floor(x*(y**3)))+random.gauss(0,1)*math.erf(random.gauss(0,1)*math.erf(x**2))",
    "f2": "random.lognormvariate(0,1)*math.asinh(random.lognormvariate(0,1)*math.floor((x**2)*(y**3))*random.lognormvariate(0,1)*abs(random.lognormvariate(0,1)*math.cos(y))+random.lognormvariate(0,1)*random.lognormvariate(0,1)*math.ceil((y**2)*x))",
    "generate": {
        "seed": 958427,
        "start": -3.141592653589793,
        "step": 0.01,
        "stop": 3.141592653589793
    },
    "plot": {
        "color": [
            -2.5634004333266436,
            ...
            2.5601010870967578
        ],
        "bgcolor": "black",
        "cmap": [
            "green",
            "white",
            "red",
            "red"
        ],
        "spot_size": 0.21,
        "projection": "rectilinear",
        "alpha": 0.1,
        "linewidth": 2.59,
        "depth": 5
    },
    "matplotlib_version": "3.2.2"
}
\end{lstlisting}